\documentclass[a4paper,12pt,twoside, onecolumn]{IEEEtran}
\usepackage{amsfonts}
\usepackage[cmex10]{amsmath}
\usepackage{graphicx}
\usepackage{multicol}
\usepackage{multirow}
\usepackage{amssymb}
\usepackage{lscape}
\usepackage[english]{babel}
\usepackage{cite}
\DeclareMathOperator{\round}{round}
\DeclareMathOperator{\expop}{\mathbb{E}}
\begin{document}
\title{An Asymptotically Efficient Backlog Estimate for Dynamic Frame Aloha}
\author{Luca~Barletta, Flaminio~Borgonovo, and~Matteo~Cesana
\thanks{L. Barletta was formerly with the Dipartimento di Elettronica,
Informazione e Bioingegneria, Politecnico
di Milano, Italy. Now he is with the Institute for Advanced Study, Technische
Universit\"at M\"unchen, Germany. (e-mail: luca.barletta@tum.de)}
\thanks{F. Borgonovo and M. Cesana are with the Dipartimento di Elettronica,
Informazione e Bioingegneria, Politecnico
di Milano, Italy. (e-mail: borgonov@elet.polimi.it; cesana@elet.polimi.it)}}
\date{}
\maketitle
\vspace{-2cm}
\begin{abstract}
In this paper we investigate backlog estimation procedures for Dynamic Frame Aloha (DFA) in Radio
Frequency Identification (RFID) environment. In particular, we address the tag identification
efficiency with any tag number $N$, including $N\rightarrow\infty$.  Although in the latter case
efficiency $e^{-1}$ is possible, none of the solution proposed in the literature has been shown to
reach such value. We analyze Schoute's backlog estimate, which is very attractive for its
simplicity, and formally show that its asymptotic efficiency is $0.311$. Leveraging the analysis,
we propose the Asymptotic Efficient backlog Estimate (AE$^2$) an improvement of the Schoute's
backlog estimate, whose efficiency reaches $e^{-1}$ asymptotically. We further show that AE$^2$
can be optimized in order to present an efficiency very close to $e^{-1}$ for practically any
value of the population size. We also evaluate the loss of efficiency when the frame size is
constrained to be a power of two, as required by RFID standards for DFA, and theoretically show
that the asymptotic efficiency becomes 0.356.
\end{abstract}


\section{Introduction}
\label{sec:intro}

Dynamic Frame Aloha (DFA) is a multiple access protocol proposed in the field of satellite
communications by Schoute  \cite{schoute}. This protocol has been rediscovered about a decade ago
for Radio Frequency Identification (RFID), an automatic identification system in which a reader
interrogates a set of tags in order to identify each one of them \cite{rfid2}. DFA and its modified
versions have then become very popular, as demonstrated by the large body of literature on the
topic, being also adopted in reference standards for ultra-high frequency RFID systems
\cite{ISO},\cite{EPC}.

In brief, Frame Aloha (FA) operates as follows: tags reply to a reader interrogation
on a slotted time
axis where slots are grouped into frames; a tag is allowed to transmit only one
packet per frame
in a randomly chosen slot. In the first frame all tags transmit, but only a part
of them avoid
collisions with other transmissions and get through. The remaining number of
tags $n$, often referred to as the
\emph{backlog}, re-transmit in the following frames until all of them succeed.

Like many protocols of the Aloha family \cite{sidi} also FA is intrinsically unstable and its
throughput is very small unless stabilizing techniques are used. This is possible if  a feedback
on the outcome of previous channel slots is available, i.e., if the reader knows whether slots
have been successfully used, not used, or collided, the latter meaning that two or more
transmission attempts have occurred. Channel feedback is then used to derive a backlog estimate
$\hat n$ and to set the frame length $r$ accordingly.

Assuming as performance figure the efficiency, defined as $\eta(N)=N/L(N)$, where $N$ the number
of tags and $L(N)$ the average number of slots needed to identify all tags, optimal conditions
arise when $\hat n=n$, i.e., the estimate is perfect, in which case the maximum efficiency is
attained by setting $r=n$ \cite{IT-12}, reaching  $e^{-1}$  as $N$ approaches infinity. For this
reason, many proposals, as discussed in the next section, set the frame length $r=\hat n$, which
surely provides optimal frame efficiency when $n \rightarrow \infty$.

Since in RFID applications the population size $N$ is an unknown constant, the initial estimate
and frame size can not be optimally set, and some frames are initially spent inefficiently waiting
for the estimate $\hat n$ to converge. The inefficiency of the initial phase constitutes an
overhead that impairs the performance with respect to the optimal case, the impairment depending
on the setting of the initial frame size and the value of $N$. This behavior is common to all
proposals appeared in the literature and, as discussed in the next section, none of those protocols
can reach efficiency $e^{-1}$  asymptotically.

In this paper, we deal with the theoretic issue represented by the asymptotic efficiency when the
initial frame length $r_0$ is finite. Although many papers have appeared proposing variations of
the basic DFA scheme, here we refer to DFA versions that are strictly compatible with the ISO and
EPC standards \cite{ISO, EPC}, not considering proposals that introduce changes in the way the
mechanism operates, sometimes borrowing mechanism from other protocol families such as Tree
protocols. The essential literature, comprising also proposals of the latter type, is given for
example in \cite{linlin, zhu2}.

We focus on the efficiency of the identification procedure. We are aware that
this figure can not consider all the details of practical implementations, i.e.,
the overhead of different commands or the fact that the slot length may change
from empty to non-empty slots; nevertheless, this figure is commonly used in the
literature to compare different protocols since it captures the essence of the
mechanisms, that is, the ability of the backlog estimate of minimizing the
identification period.

The main contribution of this paper is twofold. First we present an asymptotic analysis of
Schoute's estimate. This estimate is attractive for its simplicity, but has been often considered
inefficient and never analyzed in detail to show its potentials. Here we show that its asymptotic
efficiency, when the initial frame length is set to any arbitrary value, is $0.311$. We then
extend Schoute's estimate and propose the Asymptotic Efficient backlog Estimate (AE$^2$), an
estimate that, using the ability to restart the frame at any time even though the preceding frame
is not finished (the \emph{Frame Restart} capability of the standard), is analytically shown to
reach the asymptotic benchmark $e^{-1}$. We finally address two practical issues; namely, we show
how to tune AE$^2$ in such a way that it can guarantee efficiency close to $e^{-1}$ for any value
of $N$, and then derive the efficiency when the frame size is constrained to be a power of two, as
required by RFID standards for DFA. In the latter case we theoretically show that the asymptotic
efficiency becomes 0.356.

The paper  is organized as follows. In Sec. \ref{sec:related}, we discuss the performance of
the most relevant backlog estimates appeared in literature and state the problem. In Sec.
\ref{sec:analysis}, we provide a novel analysis of the original DFA that sheds light on its
asymptotic behavior. In Sec. \ref{sec:new}, we introduce the new estimation mechanism and
evaluate its asymptotic efficiency. In Sec. \ref{sec:practical} we address practical issues,
namely the tuning of the mechanism for best performance with finite $N$, and the evaluation of the
impairment due the constraints introduced by RFID standards. Our concluding remarks are reported
in Sec. \ref{sec:conclusions}.

\section{Background and Previous work}
\label{sec:related}

In his seminal work, Schoute proposes to estimate the backlog $n$ by counting
the number of
collided slots $c$ at the end of the previous frame. Assuming that the procedure
is able to keep a
frame size equal to the backlog $n$, the number of terminals transmitting in a
slot can be
approximated by a Poisson distribution of average $1$, such that the average
number of terminals
in a collided slot is $H= (1-e^{-1})/(1-2e^{-1})\simeq 2.39$ \cite{schoute}.
% \begin{equation*}%\label{eq:c}
% H= \frac{1-e^{-1}}{1-2e^{-1}}\simeq 2.39. \end{equation*}
The backlog estimate
is consequently
$\hat n=\round(H c)$, where $\round(x)$ is the closest integer to $x$. Moreover, Schoute shows that the average number of
slots $L(N, r)$ to
solve $N$ collisions starting with a frame of length $r$ is given by:
\begin{equation}
L(N, r)=r+\sum_{s=0}^{m}\sum_{c=0}^{\lfloor N/2\rfloor} p_{s, c}(N, r) L(N-s,
\hat r), \ \ \ \ N\ge
2, \label{eq: frame22}
\end{equation}
where $ p_{s, c}(N, r)$ is the joint probability of having $s$ successes and $c$
collisions,
\mbox{$m=\min \{N-2, r-1\}$}, and $\hat r$, equal to the estimate of the backlog $N-s$,
is function of $N,
r, s, c$. The set of equations (\ref{eq: frame22}) can be solved starting from
$L(0, r)=L(1, r)=r$.

In the past decade, many papers have faced the problem to adapt DFA to the
RFID environment.
Most of them deal with two fundamental problems: backlog estimation and frame
size determination.
Regarding the latter problem,  equation (\ref{eq: frame22}), when numerically
evaluated, shows
that it attains the minimum for $r=N$. In fact, this optimality has been
recently proven in
\cite{IT-12}, where it is also shown that the asymptotic efficiency is
$e^{-1}$. When $n$ is
unknown and there is no constraint on the frame length, Schoute and subsequent
authors have
often assumed $r=\hat n$. Recently it has been shown \cite{zhu} that setting
$r=\hat n$ is not
always optimal unless $\hat n=n$. Here, we focus on the problem of backlog
estimation and review
some of the backlog estimates proposed in literature, especially referring to
asymptotic
conditions where good estimates converge to the true value $n$.

Vogt \cite{vogt} introduces two estimation algorithms.  One is the lower bound
estimation $\hat
n=2c$, the other is the minimum distance vector based on Chebyshev's
inequality theory. The
proposed schemes are devised for a limited set of frame lengths and population
size.

In \cite{cha}, the authors consider two estimates,  the collision ratio $c/r$
and $\hat n=2.39 c$,
the same as Shoute's except for dropping the closest integer operator. Results
on optimal frame
length and collision probabilities are re-derived. The average time identification
periods of both
proposals up to 900 tags  appear practically equal.

In \cite{chen}, the value $n=\hat{n}$ that maximizes the \emph{a posteriori} probability
$\Pr(n|s,c,e)$, having
observed $s$ successes,  $c$ collisions and $e$ empty slots, is assumed as
estimate. In practice,
this proposal uses a Maximum Likelihood (ML) method since no \emph{a priori} distribution of
$N$ is given.
$\Pr(n|s,c,e)$ is obtained assuming independence among slots' outcomes. This
estimate is shown to
provide better performance than the previous ones, yielding an efficiency
$\eta=0.357$ for
$N=250$, which drops to $\eta=0.277$ for $N=50$.

%The low performance for $N=50$ shows a weakness that makes proposal unfit for a
%large tag-population size.

The drawback of this approach is its sensitivity to the initial frame size
$r_0$, which may
severely impair the performance when the initial tag population is not known, as
in most RFID
application scenarios. In fact, it is known that the ML does not
work properly
when the frame is completely filled by collisions; in this case, the ML mechanism
tends to overestimate the backlog, and the probability $\Pr(n|s,c,e)$ is maximized
by $n=\infty$.
Even if a bound $N_{\max}$ is adopted for the population size, if $r_0$ is small
and total
collision arises, the overestimation causes inefficiency; on the other side, if
$r_0$ is large,
inefficiency arises when $N$ is small.

%the initial frame size must be chosen sufficiently large, $r=128$ in the cited
%analysis, to avoid such case.
%This, however, , and
%does not work for $N\rightarrow \infty$.

This problem is common  to other proposals that are analyzed, for example in
\cite{linlin}, which
reports the evaluation of seven estimation mechanisms, including the
aforementioned ones. All of
them, except $\hat{n}=2.39c$, largely overestimate (up to ten times and more) the
backlog when the frame
is filled by collisions. The same paper also compares the efficiency of such
protocols, none of
which is shown to reach the benchmark $e^{-1}$ for large $N$. For example, the
maximum $N$
evaluated is $N=1000$, where the best mechanism considered presents an average
identification
period whose length is slightly below $300$, yielding an efficiency of about
$0.33$.

A different class of estimates is given by the Bayesian estimate in \cite{flor2}
that evaluates
the a posteriori probability distribution of the original population size $N$,
conditioned to all
the past observations, starting from the a priori distribution of the number of
transmitting tags.
Here the problem is that such a priori distribution is not known, and can not be
hypothesized with
an unbounded $N$.

The above discussion shows the inadequacy of the proposals when the population
size does not have a
known distribution or the population size is large. However, to better sustain
this inadequacy, we
have re-evaluated the performance of the two most relevant schemes, namely the
Schoute's proposal
and the Bayesian one.

\begin{figure}
\centering \includegraphics[width=\columnwidth]{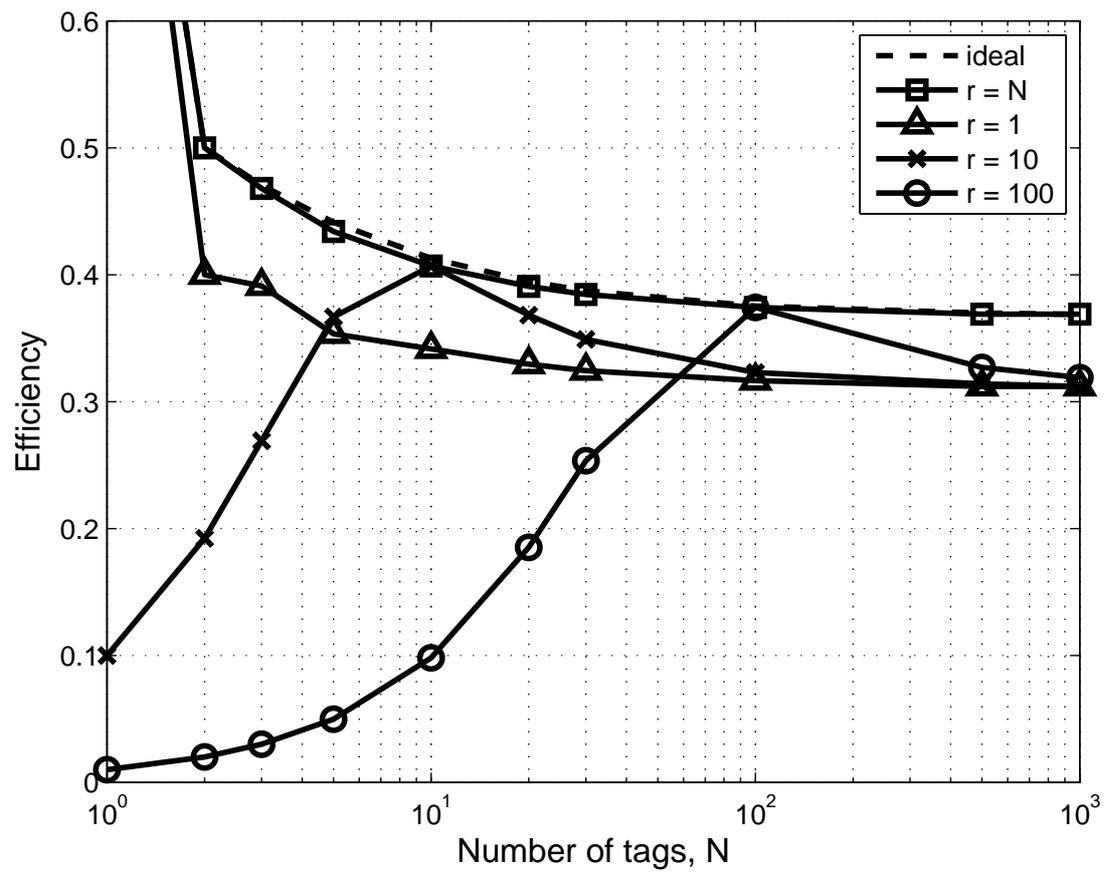}
\caption{\emph{Efficiency
versus the initial number of tags N of Schoute's DFA
mechanism for different
values of the initial frame length $r$.}} \label{fig0} \vspace{-0.5cm}
\end{figure}

In Fig. \ref{fig0}  the efficiency for different values of $N$ with initial
frame length $r_0=N$,
$r_0=1$,  $r_0=10$  and $r_0=100$ is shown. Values up to $N=30$ have been
evaluated using
(\ref{eq: frame22}), whereas values for $N=500$ and $N=1000$ have been obtained
by simulating the
algorithm. To allow comparisons we have also reported the performance with a
perfect estimate
$\hat n=n$ (dashed line), that represents a benchmark for all estimation
mechanisms. We have also
reported the case where only the estimate of the first frame is perfect, i.e.
when the first frame
length is set to $N$. The comparison of the latter cases shows that Shoute's
mechanism is able to
track the backlog quite well if compared to the perfect estimate case,
asymptotically reaching the
best possible efficiency $e^{-1}$. In all the other cases, Schoute's estimates
suffer the mismatch
between $N$ and the initial frame length $r$, and the efficiency degrades
monotonically when $N$
increases beyond $r$, indicating the existence of a possible asymptote well
below $e^{-1}$. The
reasons for such behavior are analyzed next.

\begin{figure}
\centering \includegraphics[width=\columnwidth]{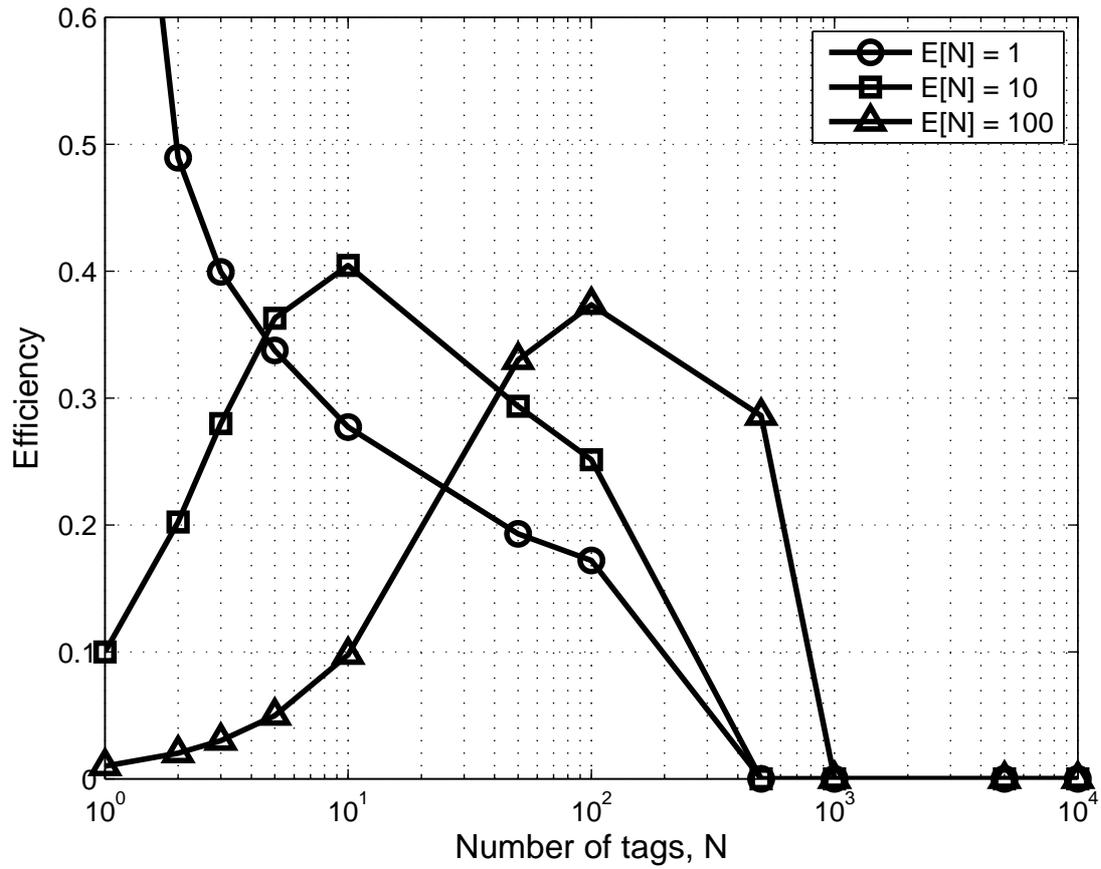}
\caption{\emph{Efficiency
of DFA with Bayes estimate and Poisson population-size distribution.}}
\label{fig9}
% \vspace{-0.9cm}
\end{figure}

\begin{figure}
\centering \includegraphics[width=\columnwidth]{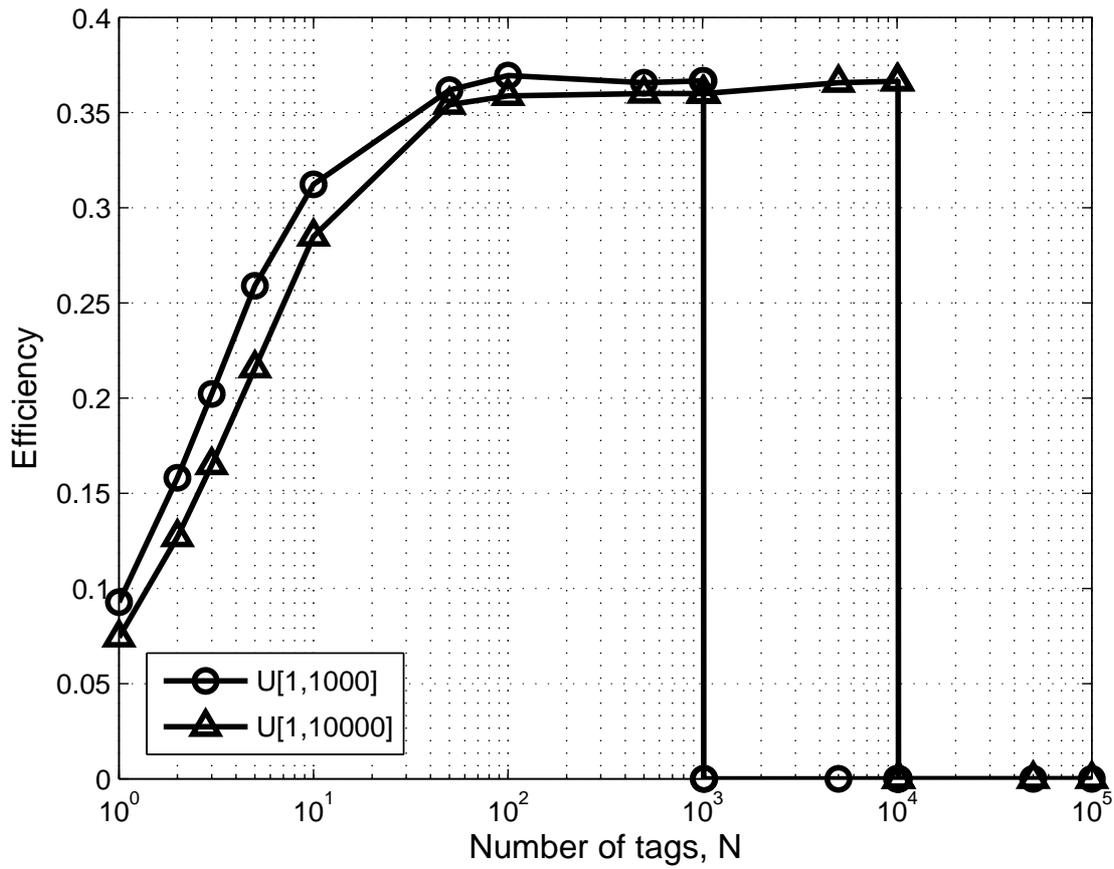}
\caption{\emph{Efficiency
of DFA with Bayes estimate and Uniform population-size distribution.}}
\label{fig10}
% \vspace{-0.9cm}
\end{figure}

Figure \ref{fig9}  shows the efficiency of the Bayesian method when the
population size
distribution is Poisson with different averages. We see that the efficiency is
optimal only when
$N$ lies around the average. In all cases, when $N$ increases beyond some
maximum value the
efficiency drops to zero, meaning that the estimate converges to a constant
value well below the
actual value, so that successes are very rare. In Fig. \ref{fig10} the
performance with a
uniform population size shows a similar behavior.

None of the two methods just shown seems to behave well asymptotically; however,
the Schoute's
method is by far the simplest of the two and presents characteristics, namely
the ability to
remain locked to the true value, that suggest that it can be improved to reach
maximum
asymptotic efficiency. To this purpose, we present in the next section an
asymptotic analysis
that suggests how to reach the target.

\section{Asymptotic analysis of Schoute's method}\label{sec:analysis}
In the following analysis we assume that $N$ is very large, since we are
interested in
investigating the efficiency for $N \rightarrow \infty$. To facilitate the
reader, we proceed in
steps. In the remainder of the paper lower case letters represent random
variables, whereas
calligraphic upper cases represent averages.

{\bf Step 1}. Here we derive recursive formulas for the backlog.  We initially
assume that the
frame size $r_i$, and the backlog $n_i$ are so large that the number of
transmissions in a slot
can be approximated by a Poisson distribution with average $n_i/r_i$. This
allows to evaluate the
probability of an empty, successful and collided slot as
\[
p_e=e^{-n_i/r_i}; \qquad p_s=\frac{n_i}{r_i} e^{-n_i/r_i}; \qquad
p_c=1-\frac{n_i}{r_i}
e^{-n_i/r_i}-e^{-n_i/r_i}. \]

%\begin{equation}\label{eq:pr}
%\begin{array}{l}  p_e=e^{-n_i/r_i} \\
%p_s=\frac{n_i}{r_i} e^{-n_i/r_i} \\
%p_c=1-\frac{n_i}{r_i} e^{-n_i/r_i}-e^{-n_i/r_i}.
%\end{array}
%\end{equation}

We note that relations above also hold when starting with small $r$, because in
this case, being
$N-i$ always very large, every slot is collided with probability one.  In
Appendix A we show that,
in the conditions assumed, the ratio $k_i=n_i/r_i$ can be safely replaced by the
ratio of the
respective averages ${\cal K}_i={\cal N}_i/{\cal R}_{i}$, which is the traffic
per slot. With this
substitution the probabilities above are denoted by ${\cal P}_e, {\cal P}_s,
{\cal P}_{c}.$ This
means that the average number of collisions and the average backlog size can be
expressed as
\begin{equation} \label{eq:sl0}
 {\cal C}_i={\cal R}_i {\cal P}_c, \qquad {\cal N}_{i+1}={\cal N}_{i}(1-{\cal P}_s).
\end{equation}

The frame length evolves with law $ r_{i+1} = \round (H c_i)$,  so that
\begin{equation}\label{eq:sl1a}
{\cal R}_{i+1} = \expop\left\{\round (H c_i)\right\},
\end{equation}
where $\expop\{\cdot\}$ is the expectation operator.
Equations (\ref{eq:sl0}) and (\ref{eq:sl1a}) form a recursion
that provides
sequences $\{{\cal R}_{i}\}$ and $\{{\cal N}_{i}\}$ that determine the
efficiency. Unfortunately
the rounding operation in (\ref{eq:sl1a}) makes their analysis practically
unfeasible.

{\bf Step 2}. When $c_i$ is large, exploiting the limit $\lim_{x\rightarrow
\infty} \round(x)/x =
1$, we can replace  $\round(H c_i)$ in (\ref{eq:sl1a}) with $ H c_i$,
obtaining
\begin{equation}\label{eq:sl1aa}
 {\cal R}_{i+1} =\expop\left\{\round(H c_i)\right\}\simeq H
\expop\left\{c_i\right\}= H
 {\cal C}_i.
\end{equation}
If we use (\ref{eq:sl1aa}) in the recursion in place of (\ref{eq:sl1a}) we get
sequences
\begin{equation} R_{i+1}=H R_{i}\left(1-K_i e^{-K_i}-e^{-K_i}\right),
\label{eq:slots9}
\end{equation}
\begin{equation}
N_{i+1}=N_{i}\left(1-e^{-K_i}\right), \label{eq:slots9a}
\end{equation}
\begin{equation}\label{eq: k_inf}
K_{i+1}=K_i\ \frac{1}{H}\frac{1-e^{-K_i}}{1-K_ie^{-K_i}-e^{-K_i}},
\end{equation}
that correspond, respectively, to sequences $\{R_i\}$, $\{N_i\}$, and $\{K_i\}$.
 Later on we prove
that replacing  $\{{\cal R}_{i}\}$, $\{{\cal N}_{i}\}$, and $\{{\cal K}_{i}\}$
with the above
sequences has no effect on the evaluation of the asymptotic performance. Also we
prove that this
holds even for finite values of the initial frame size $r_0$. In practice, we
find that sequence
$\{R_i\}$ approximates fairly well sequence $\{{\cal R}_{i}\}$, even for
moderate values of $N$,
and this allows recurrence (\ref{eq: k_inf}) to be used in practice to evaluate
the performance.

\begin{figure}[t]
\centering \includegraphics[width=\columnwidth]{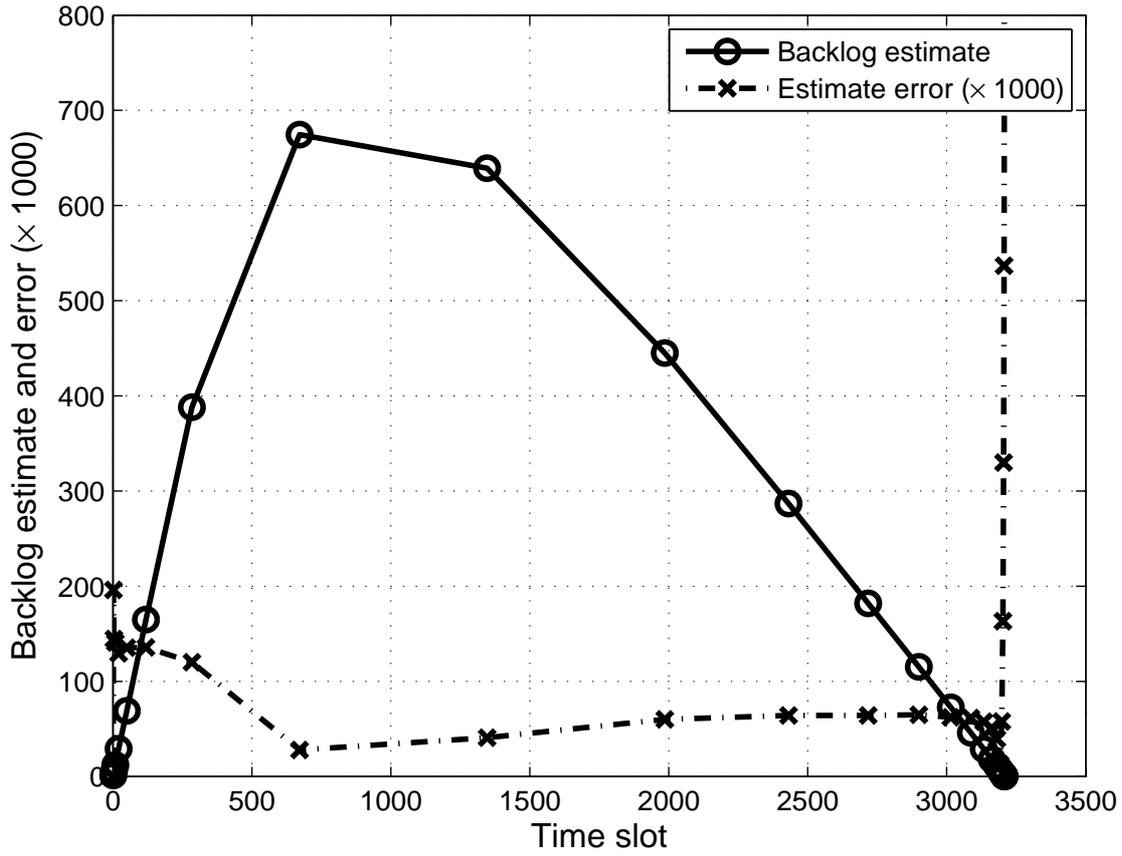}
\caption{Average Schoute's backlog estimate $\hat {\cal N}_{i}$ at the end of
the frames versus time slot ($N=1000$, $r=1$). The dash-dotted line represents
the relative error $\times 10^3$ with respect to the analytical values $N_i$.}
\label{fig5}
\end{figure}

As an example, Fig. \ref{fig5} shows sequence $\{ {\cal N}_{i}\}$ derived by
averaging $10^4$
simulation samples in the case $N=10^3$ and $r_0=r=1$. We can clearly see a
first phase where the
estimate increases in order to converge to the true value $N=10^3$. In a second
phase, optimal
conditions are met, collisions are solved and the backlog decreases steadily to
reach zero at
about the 25-th iteration. We explicitly note that in the descending phase the
rate of descent is
$e^{-1}$, showing that Schoute's algorithm is capable to correctly track the
backlog and to solve
contentions in the most efficient way. Figure \ref{fig5} also shows the relative
error sequence
$\{({\cal N}_{i}-N_{i})/{\cal N}_{i}\}$ multiplied by $10^3$ (dash-dotted line). The
error is always
very small except at the end of the process, where ${\cal N}_i$ becomes small
and ignoring the
rounding effect is no longer appropriate. However, this error has no effect on
the efficiency
since it occurs for a small period of time, negligible when compared to the
entire collision
resolution length.

\begin{figure}[t]
\centering \includegraphics[width=\columnwidth]{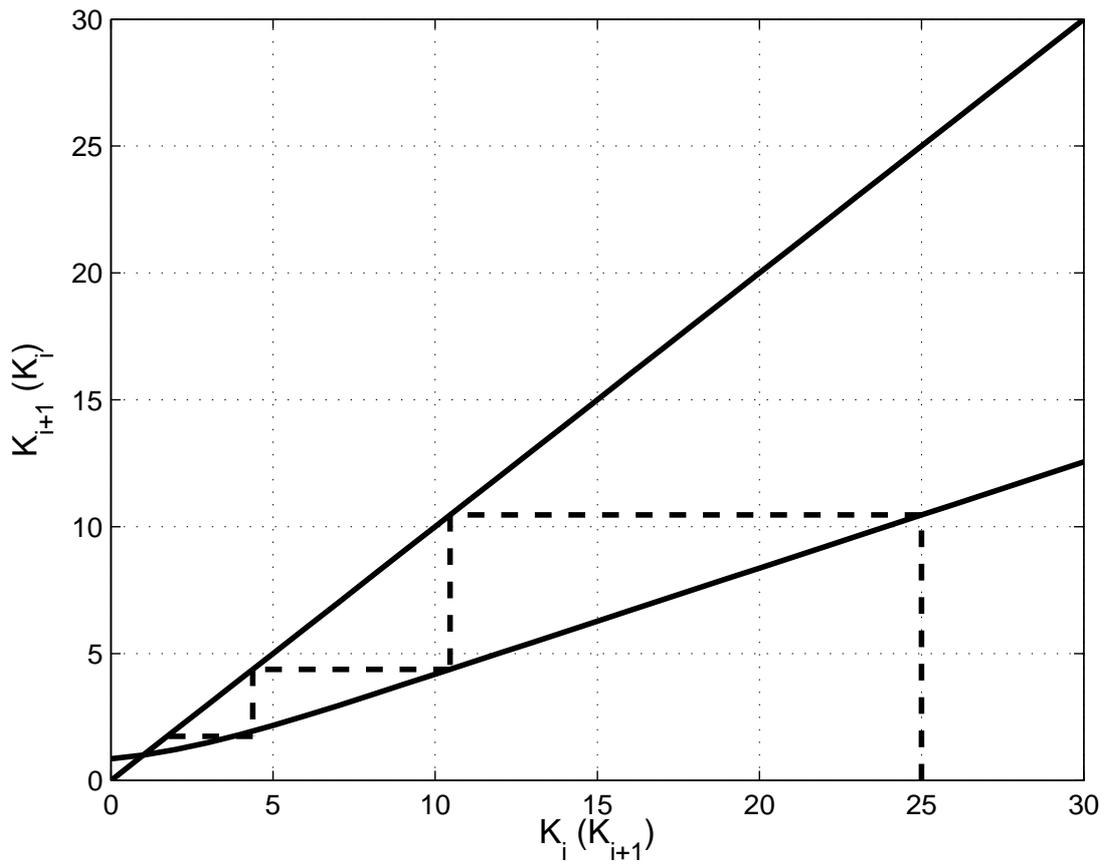}
\caption{Representation of the trajectory of the sequence  $\{K_i\}$.}
\label{fig3}
%\vspace{-0.9cm}
\end{figure}

{\bf Step 3}. The evolution of the entire process is represented by recurrence
(\ref{eq: k_inf})
that depicts the evolution of average traffic $K_i$. This can be represented by
the dashed
trajectory in Fig. \ref{fig3}. This figure also shows that the evolution of
the process is
asymptotically stable since recurrence (\ref{eq: k_inf}) leads to the fixed
point in $K_i=1$. This
point is also a point of optimality because in here we attain the optimal
condition $r_i=n_i$ that
provides maximum throughput.

When the starting point  in (\ref{eq: k_inf}) is $K=K_0=1$,  the collision
resolution process
proceeds with a correct backlog estimate, yielding $K_i=1$, for all subsequent
$i$ and we have
\begin{equation}\label{eq: l1_inf}
R_{i+1}=(1-e^{-1})R_i, \qquad i\ge 0.
\end{equation}
The solution of  the recurrence (\ref{eq: l1_inf}) is $R_{i}=(1-e^{-1})^{i}N$,
for $i\ge 0$. The
total number of slot in this resolution phase is $ L(N)=\sum_{i=0}^{\infty}R_i
=Ne$, yielding an
asymptotic throughput $N/L(N)=e^{-1}$. When $K =N/r >1$, the length of the
entire procedure can be
evaluated as $L(K)=\sum_{i=0}^\infty R_i=  r \sum_{i=0}^\infty a_i$,
with $R_0=r$, where sequence $\{a_i=R_i/r\}$ is always the same, for a given
$K$, whichever $r$
is. Therefore, the efficiency is evaluated  as $N/L=K/\sum_{i=0}^\infty a_i$ and only depends on $K=N/r$.

{\bf Step 4}. Here we show that for large values of initial traffic $K$ the
dependence of the
efficiency on $K$ is negligible. Since the protocol always starts with a finite
$r$, large $N$
means large $K$, so we attain practically the same efficiency whichever the
initial frame length
$r$ is.

\begin{figure}
\centering  \includegraphics[width=\columnwidth]{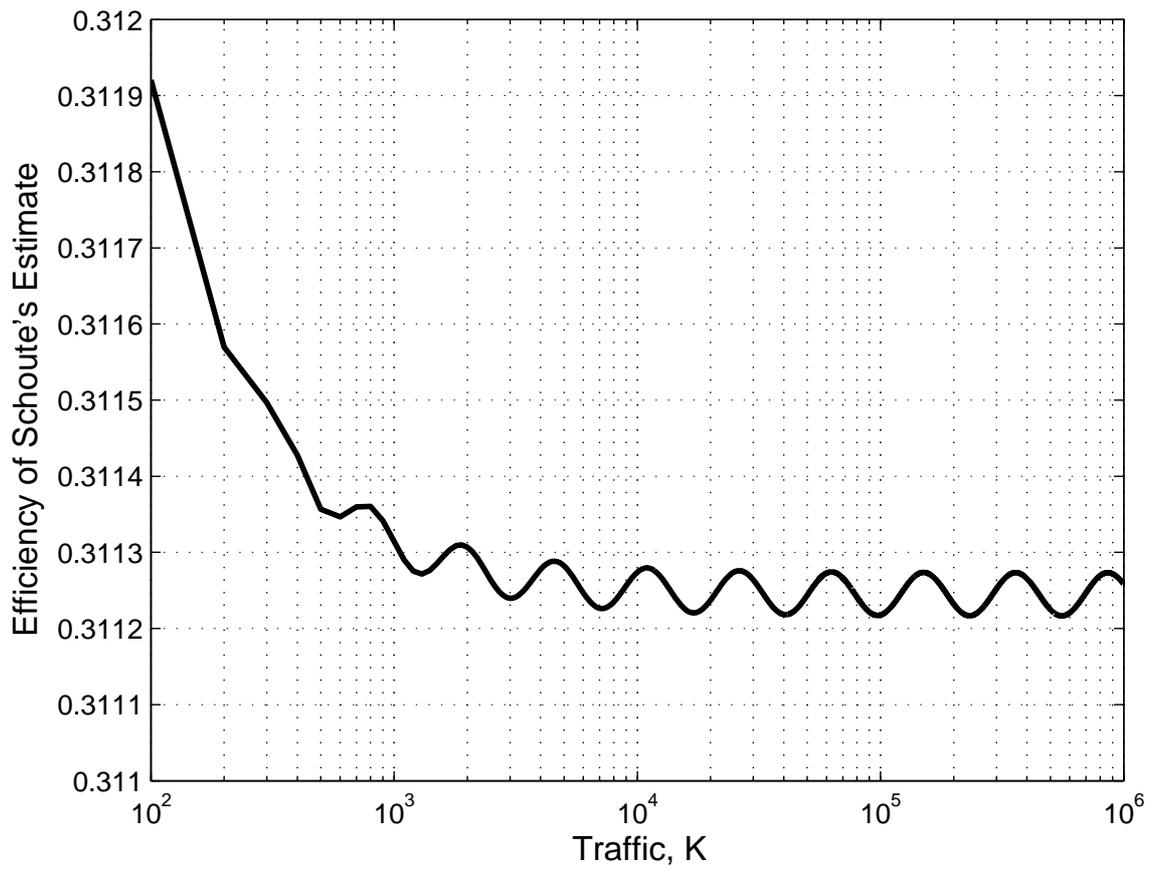}
\caption{Efficiency of Schoute's backlog estimate versus initial traffic $K$.}
\label{fig4} %
\end{figure}

As an example, in Fig. \ref{fig4} we have reported the efficiency $N/L(N)$,
evaluated through
(\ref{eq:slots9}) and (\ref{eq: k_inf}), for different values of traffic $K$.
Starting from $K=1$
the efficiency at first decreases as $K$ increases until about $K=500$ where it
begins to oscillate
without reaching an asymptote, around a mean value of $0.31125$, with a period
that
increases geometrically with $H$.

To attain some insight on the asymptotic behavior, during the solving process
we consider three
phases. The first phase, the \emph{approaching phase},  starts at frame $0$ with
infinite traffic and ends
at frame $u$ where $u$ is chosen in such a way that the traffic  $K_u$ is finite
and practically no
successes occur up to frame $u$; as an example we may arbitrarily assume $u$
such as $K_u\ge 10$.
Although in this way $K_u$ and $u$ appear arbitrarily defined, we show below
that this has no
effect on the evaluation of the efficiency, as, in fact, the initial traffic $K$
has no effect.
The assumed definition for $u$ assures that $u \rightarrow \infty$ as $N
\rightarrow \infty$ and
$R_u=N/K_u$.

The second phase, the \emph{convergence phase}, starts at frame $u+1$ and ends at frame
$u+v$ such that
$K_{u+v}\simeq 1$. At this point the third phase, the \emph{tracking phase}, begins
where tags are solved
with efficiency $e^{-1}$. Denoting by $L',L''$, and $L'''$ the length of the
three phases
respectively, the efficiency is evaluated as
$ N/L(N)= N/(L'+L''+L''')$.

With high values of $K=N/r$, in the first phase the frame length increases
deterministically with
law $ R_i=r H^{i}$, for $i \ge 0$.
The average number of slots up to frame $u$ where the first phase ends is
\begin{equation*}%\label{eq:fase1}
L'=\sum_ {i=0}^u R_i=  r\frac{H^{u+1}-1}{H-1}\simeq  \frac{H}{H-1}R_u.
\end{equation*}
Replacing $R_u=N/K_u$, the average length of the first phase becomes
\begin{equation*}%\label{eq:fase3}
L'= \frac{H}{H-1} \frac{N}{K_u}=N A(K_u).
\end{equation*}
where $A(K_u)$ is the proportionality constant where we have made explicit the
dependence on $K_u$.
The second phase starts at frame $u+1$, when $K_{u}$ is such that
the collision probability is practically one,  and ends at frame
$u+v$ when $K_{u+v}\simeq 1$. Equation (\ref{eq:slots9}) can be
used to evaluate the length of phase two by the following sum over
a finite number of terms:
\begin{equation*}%\label{eq:fase4}
L''=\sum_ {j=1}^v R_{u+j}= R_u \sum_ {j=1}^v \alpha_j= N B(K_u),
\end{equation*}
where terms $\alpha_j$ are all finite and, again, where $B(K_u)$ is the
proportionality constant where we have made explicit the dependance on $K_u$.
The average backlog size at the end of the second phase can be
evaluated by (\ref{eq:slots9a}) as
\begin{equation*}%\label{eq:fase5}
N''=N_{u+v}= N_u \prod_ {j=1}^v \left(1-e^{-K_{u+j}}\right)= N C,
\end{equation*}
where we have exploited the fact that $N_u=N$. Coefficient $C$ does not depend
on $K_u$, since in
frame $u+1$ we still have all collisions ($e^{-K_{u+1}} \simeq 0 $).

The third phase presents  efficiency  $e^{-1}$ and its average
length is $L'''=N'' e=NC e$. The efficiency with very large $N$ is
then
\begin{equation}\label{eq:fase7}
 \frac{N}{L(N)}= \frac{N}{L'+L''+N''e}=
\frac{1}{A+B+C e}.
\end{equation}
We note that (\ref{eq:fase7}) does not depend on the choice of
$v$, once the condition $K_{u+v} \simeq 1$ is assured. If we
replace $v$ by $v+1$, coefficient $A$ is not affected, and also
term $B+C e$ is  not affected. In fact, $B$ is augmented by the
term $R_{u+v+1}$ which, by (\ref{eq:slots9})  with  $K_{u+v+1}
\simeq 1$, is equal to
\begin{equation}\label{eq:fase8} R_{u+v+1}=N_{u+v}(1-e^{-1}). \end{equation}
On the other side, term $C e$ is diminished by $(N_{u+v}-N_{u+v+1})e= N_{u+v}(1-e^{-1})$, that is equal to term (\ref{eq:fase8}). Nevertheless, efficiency
(\ref{eq:fase7}) does depend on
the choice of $K_u$, through coefficients $A$ and $B$. However, if we replace
$K_u$, chosen as
suggested above, with $K_u \cdot H$, efficiency (\ref{eq:fase7}) does not
change because this only
implies the shifting of term $R_u$ from term $A$ to term $B$. Therefore, the
efficiency is
periodic in a logarithmic scale and all the asymptotic amplitudes of the
oscillations in Fig.
\ref{fig4} can be obtained by replacing $K_u$ with any value $K'$ in the range
$(K_u, H K_u)$.

Table \ref{tabfr4b} shows the efficiency attained by (\ref{eq:fase7}) for
different values of
$K_u$ chosen in the range $(20,20 H)$. As we can see, the values fit very well
to those shown in
Fig. \ref{fig4} and, for all practical purposes, the asymptotic efficiency
can be assumed
equal to $0.311$.

{\bf Step 5}. Now we show that replacing $L', L''$, and $L'''$, in the limit
$r\rightarrow \infty$,
with $\mathcal{L'},\mathcal{L''}$, and $\mathcal{L'''}$, in which the rounding
operation is taken
into account, does not change the results provided that the initial frame
length is still $r$. In Appendix B %\ref{sec:appendix_B}
 we show that
\[
\lim_{r\rightarrow\infty}\frac{\mathcal{L'}(N)}{L'(N)}=\frac{\sum_{i=0}^{\infty}
\mathcal{R}_i}{\sum_{i=0}^{\infty}R_i}=1.
\]
We also have $\lim_{r\rightarrow\infty} \mathcal{L''}(N)/L''(N)=1$, because the second phase is composed of a finite number $v$ of frames, each of
them so large that
the rounding effect is negligible. What shown also implies that at the end of
the second phase we
have $\lim_{r\rightarrow\infty}\mathcal{N}_{i}/N_{i}=1$, and,
therefore, since those tags are solved with efficiency $e^{-1}$,
also for the length of the third phase we have $\lim_{r\rightarrow\infty} \mathcal{L'''}(N)/L'''(N)=1$.

{\bf Step 6}. If $r$ is small and (\ref{eq:sl1aa})  can not be assumed, the
first phase is split
into two sub-phases in which the second sub-phase starts at frame-index $x$ such
that, from this
frame onward, the rounding operation in (\ref{eq:sl1a}) can be disregarded. Index
$x$ is finite and
the length of the first sub-phase does not depend on $N$, whereas the  length of
the second
sub-phase and of the other phases is proportional to $N$. Therefore, as $N
\rightarrow \infty$,
the length of the first sub-phase vanishes and the asymptotic efficiency
remains approximately
$0.311$ even with small $r$.

\section{An Asymptotically Efficient Estimation Procedure}
\label{sec:new}

The analysis of Schoute's estimate carried out in the previous section has shown
that the
reduction of the asymptotic efficiency with respect to the theoretical value
$e^{-1}$, when
starting with a finite estimate, is not due to an intrinsic inefficiency of the
estimate, but
rather to the phase in which traffic $K$ converges to  $1$, that is the
convergence phase composed
of $L'$ and $L''$, whose length increases linearly with $N$. In particular, the
length of this phase
increases linearly with $N$ because the frame length increases exponentially as
$H^i$, and this,
from the overhead point of view, is a complete waste of time, since in this
phase almost no success
occurs. On the other side, the frame increase is needed to reduce the traffic
per slot and get
locked to the optimal point $K=1$. To get a good estimate of traffic $K$, we
need not to explore
the entire frame or, in another view, we need not to let all tags transmit in
the frame;
therefore, during the approaching phase toward $K=1$ the frame can be shorter
and provide a convergence phase with an average length $L'+L''$ such that
\begin{equation}\label{eq:pre0}\lim_{N \rightarrow \infty } \frac {L'+L''}{N}=0.
\end{equation}

A way to reduce the number of tags transmitting in the frame is to specify at
the beginning of the
frame the transmission probability, together with the frame length. An
alternative way, that is
entirely compatible with the EPC standard, is to let all tags chose a slot in
the frame, as in
normal operation, but re-starting a new frame before the exploration of the
entire frame is
completed. Therefore, if we call {\em virtual frame}, of length $z$, the frame
in which all tags
select a slot for transmission, and {\em real frame}, of length $r \le z$, the
frame that has been
explored when the frame is re-started, the traffic per slot $n/z$ is determined
by  the virtual
frame, but only the real frame is observed and used to determine $\hat{ n}$.

The estimation procedure we present in this section, the AE$^2$, adopts a  virtual frame whose length $z_i$ is set equal to the backlog
estimate $\hat
n_i$, as it happens in the Schoute's mechanism. Furthermore, the  real frame
length $r_i$ is set
so that, in the convergence phase, it increases with index $i$ far less than the
virtual frame
length. As in DFA, backlogged terminals choose a slot in the virtual frame
length $z_i$ and
transmit in it only if the chosen slot belongs also to the real frame, i.e., if
the \emph{Frame Restart}
command has not arrived yet. By setting a suitable law for $r_i$, the length of the first two phases can be easily forced to obey (\ref{eq:pre0}).

The backlog estimate is updated as follows
\begin{equation}  \label{eq:pre2} \left.\begin{array}{ll}
z_{i+1}&=\hat{n}_{i+1}=\round \displaystyle  \left( H_i
\frac{z_i}{r_i}c_i\right), \qquad c_i >0, \\
z_{i+1}&=z_i-s_i, \qquad c_i=0. \end{array}\right.
 \end{equation}
Update (\ref{eq:pre2}) is a variation of Schoute's algorithm. It uses the
number of collided
slots $c_i$ in the real frame to get an estimate, $(z_i/r_i)c_i$, of the number
of collided slots
in the virtual frame, multiplied by the factor $H_i$. When $c_i=0$, however, no
estimate can be
inferred by the observation, and, therefore, the estimate is assumed identical
to the one in the
previous frame diminished by the observed number of successes $s_i$. In
Schoute's work, where the
algorithm is supposed to operate in $K=1$, we have $H_i=H$, for all $i$; in our
case, however,
such a setting cannot guarantee the convergence to $K_i=1$. This issue is
discussed in Sec. \ref{sec:AE}.

As for the length of the real frame $r_i$, we asymptotically use the law
\begin{equation}\label{eq:pre4}
r_{i}= \min \{ \round((i+1)^b),
{z_i}\}, \end{equation}
with $b>0$. In (\ref{eq:pre4}), with large $N$, with the exception of the first few slots,
the real frame size
increases, at first, as $(i+1)^b$, and later, when $z_i$ stabilizes and $i$ is
such that
$\round((i+1)^b)> z_{i}$, the real frame coincides with the virtual one
and the procedure becomes
the classic DFA.

The proposed procedure resembles in some aspects the one in
\cite{kodia}, where the
traffic is decoupled from the frame length by the introduction of a frame
transmission probability
less than one. However, unlike our proposal where estimation takes place within
the identification
phase, in \cite{kodia} the identification phase is preceded by an estimation
phase, introduced \emph{ad
hoc}, whose length increases as $ \log N$ and depends on the length of the
estimate confidence
interval, that must be made quite small, as no estimation is operated during the
identification
phase. The latter characteristic can jeopardize the procedure since no certainty
exists to
identify all tags.

\subsection{Asymptotic Analysis of AE$^2$}\label{sec:AE}

The analysis here presented is much the same as the one  presented in Sec.
\ref{sec:analysis}.
Therefore we limit our explanation to parts that differ. Adopting the same
assumptions  used in
Sec. \ref{sec:analysis} we can write the recursions corresponding to
(\ref{eq:slots9})-(\ref{eq: k_inf}) as
 \begin{equation}\label{eq: 2l_inf}
Z_{i+1}=Z_i\ H_i\left(1-K_i e^{-K_i}-e^{-K_i}\right)
\end{equation}
\begin{equation*}%\label{eq:2rec2}
  N_{i+1} =  N_i\left(1-\frac{R_i}{Z_i}e^{-K_i}\right)
\end{equation*}
\begin{equation}\label{eq: 2k_inf}
K_{i+1}=K_i\ \frac{1}{H_i}\frac{\displaystyle
1-\frac{R_i}{Z_i}e^{-K_i}}{1-K_ie^{-K_i}-e^{-K_i}}.
\end{equation}
The key recursion (\ref{eq: 2k_inf}) is different from (\ref{eq: k_inf})  since
now it also
depends on $Z_i$ which complicates the matter. Since, for an efficient
estimation we want $K_i$ to
converge to $1$, sequence $\{H_i\}$ must be chosen as
\begin{equation}\label{eq: 2seq}
H_i = \frac{\displaystyle 1-\frac{R_i}{Z_i}e^{-1}}{1-2e^{-1}}.
\end{equation}
Recursion (\ref{eq: 2k_inf}) is stable because it presents a
unique fixed point in $K=1$ and we have
\[
-1<\left.\frac{\partial}{\partial K}\left\{K\ \frac{1-2e^{-1}}{1-Be^{-1}}\
\frac{1-Be^{-K}}{1-Ke^{-K}-e^{-K}}\right\}\right|_{K=1}<1,
\]
for all $B\in (0,1].$ Although values (\ref{eq: 2seq}) could be evaluated a
priori, in
practice we can assume
\[%begin{equation}\label{eq: 2seqb}
H_i =  \frac{\displaystyle 1-\frac{r_i}{z_i}e^{-1}}{1-2e^{-1}}.
\]%end{equation}

\begin{figure}[t]
\centering  \includegraphics[width=\columnwidth]{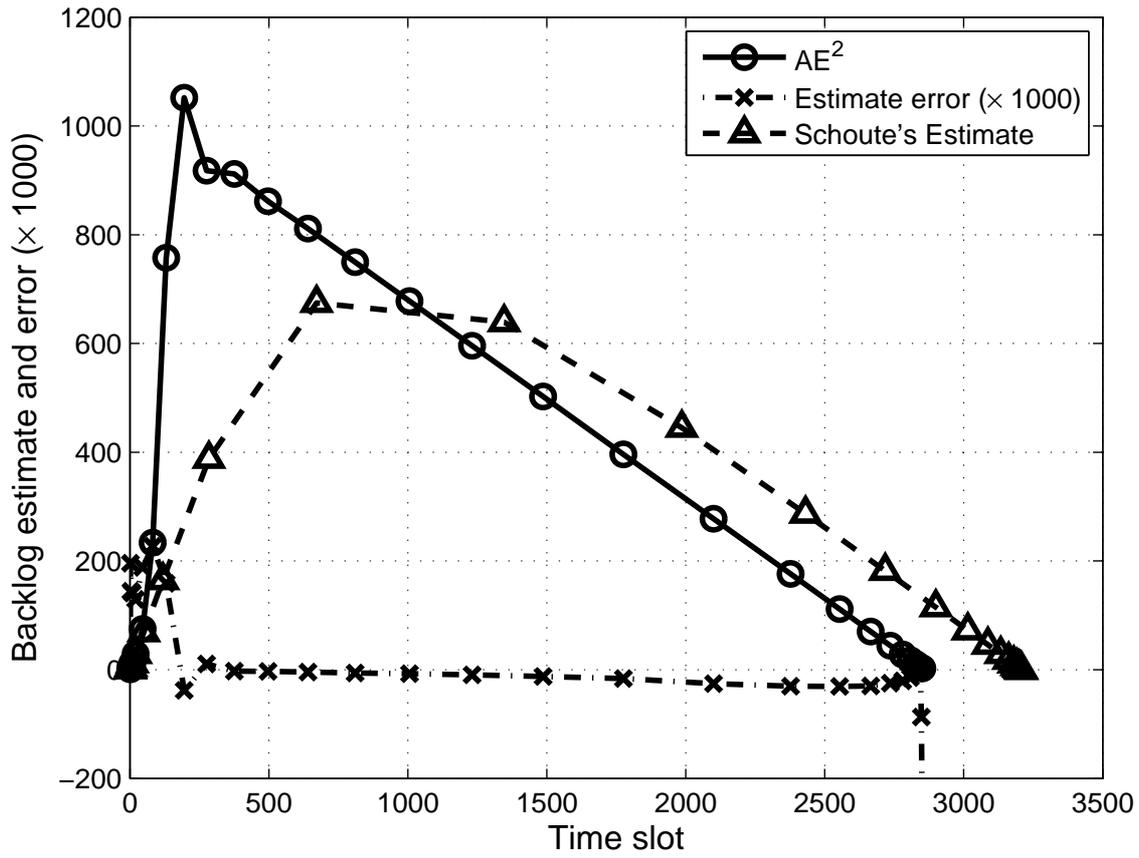}
\caption{Average backlog estimate $\hat
{\cal N}_{i}$ at the end of frames versus time slot, for the AE$^2$ algorithm
($N=1000$, $r=1$, $b=2$). The dash-dotted line
represents the relative error $\times 10^3$ with respect to the analytical
values $N_i$.}
\label{fig6}
\end{figure}

Figure \ref{fig6} validates the analysis so far carried out. In fact, it
compares the results the
analysis produces in terms of sequence  $\{{\cal N}_i\}$ with exact values
attained averaging
$10^4$ simulation samples, in the case $N=10^3$ and $r=1$. Again, the dash-dotted
line represents the
relative error multiplied by $10^3$, still very small. For comparison purposes
we have also
reported the curve in Fig. \ref{fig5} that refers to Schoute's algorithm.
We clearly see the
advantage of AE$^2$: the estimate $ \hat {\cal N}_i$  at first rises sharply
reaching $N$ with some
overshoot, higher and sooner with respect to Schoute's case. Right after the
estimate begins a
steady decline with rate $ e^{-1}$.

\begin{figure}
\centering \includegraphics[width=\columnwidth]{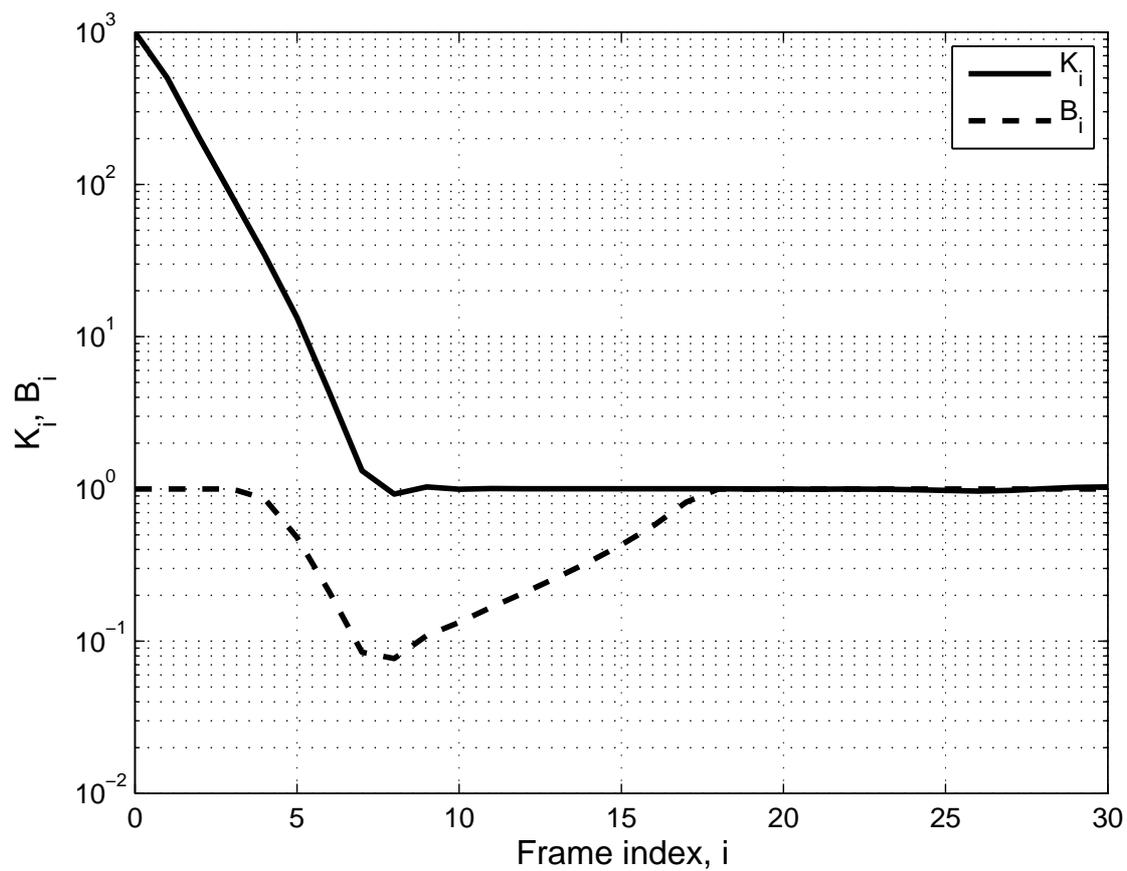}
\caption{Sequences $\{K_{i}\}$ and $\{B_i\}$ versus the frame index $i$
($N=1000$, $r=1$, $b=2$).} \label{fig7}
\end{figure}

What stated above is confirmed in Fig. \ref{fig7} where we have reported
sequences $\{K_i\}$ and
$\{B_i\}=\{R_i/Z_i\}$. The former shows the convergence of the estimate in
$K=1$, while the latter
reports the convergence of the real frame $R_i$ to the virtual frame $Z_i$. The
protocol starts
with  $r_0=z_0$, $ e^{-K_i}\simeq0$ and subsequently we have $H\simeq 2.39$ as
in Schoute's, which
yields  $z_{1}=r_1=2$. Condition $r_i=z_i$ is maintained up to $i=3$  and then
becomes $z_i>r_i$.
$B_i$ decreases and when $r_i\ll z_i$ we have $H_i=H'\simeq 1/(1-2e^{-1})\simeq
3.78$, and
$z_{i+1}=H' z_i$, reducing the traffic more quickly than in Schoute's and speeding
up the
convergence phase, which is further reduced in time because the real frame is by
far shorter.
$B_i$ reaches a minimum when $K_i$ reaches one. At this point the real frame is
so short that the
collision solved are still very few. Beyond this point the protocol solves
collisions with
efficiency $ e^{-1}$, $z_i$ decreases and $r_i$ increases until condition
$r_i=z_i$ is reached
again  and never abandoned. From this point onward the backlog is solved
exactly as in Schoute's
algorithm.

It is worth noting that the recursion in $B$ does not get into its fixed point
$B=0$. In fact,
once $K=1$ is reached, by (\ref{eq: 2l_inf}) we can write
$ B_{i+1}/B_{i}=R_{i+1}/R_i (1-B_{i}e^{-1})^{-1} > 1$.

% %\begin{figure} \centering \epsfxsize=1 \columnwidth \epsffile{fig8.eps}
% \vspace{-0.6cm}
% %\caption{\emph{Efficiency of AE$^2$ algorithm versus the frame index $i$ for
% different values of
% % parameter $b$. The dashed line represents the efficiency of the modified
% AE$^2$ algorithm.}}
% % \label{fig8} \vspace{-0.5cm}
% %  \end{figure}

Now we prove that the efficiency of AE$^2$ equals $e^{-1}$. The efficiency can
be evaluated by
writing, as in Sec. \ref{sec:analysis},
\begin{equation}\label{efficiency}
\eta= \lim_{N\rightarrow \infty}\frac{N}{L^{'}(N)+L^{''}(N)+N^{''}e},
\end{equation}
where $L^{'}(N)$ is the average number of slots of the first phase in which
there are no
successes, $L^{''}(N)$ is the average number of slots of the second phase in
which the estimate of
the backlog converges to the actual backlog. As it has been already observed, in
the first phase
we have $H_i=H'$, so that we have $Z_i =r (H')^{i}$,  and assuming, as in
the earlier
analysis, that the first phase ends at frame $u$, where $Z_u=N/K_u$, solving the
expression of
$Z_u$ we get
\begin{equation}\label{efficiencyb}
 u(N) = \log_{H'}\frac{N}{rK_u}.
\end{equation}
The length of this phase can be bounded as
\begin{equation}
L^{'}(N)\le \sum_{i=1}^{u(N)}i^{b+1}\le\int_0^{u(N)+1}i^{b+1}\
di=\frac{(u(N)+1)^{b+2}}{b+2}\le  \frac{\left(1+\log_{H'}\frac{N}{r
K_u}\right)^{b+2}}{b+2},\label{efficiencyc}
\end{equation}
where last inequality follows by (\ref{efficiencyb}),
therefore we have $\lim_{N \rightarrow \infty} L^{'}(N)/N=0$. The overhead of the second phase can be rewritten as
\begin{equation}\label{efficiencyk}
L^{''}(N) = (N-N^{''})(e-\epsilon),\qquad 0<\epsilon<e,
\end{equation}
where $N^{''}$ is the backlog size at the end of the second phase. At the end of
the first phase we have
\[
B_u\triangleq\frac{R_u}{Z_u}\approx\frac {K_u
u^b}{N}=\frac{K_u(\log_{H'}N-\log_{H'}r K_u)^b}{N},
\]
which implies $ \lim_{N\rightarrow \infty}B_u=0$.
In the second phase a few frames, $v$, are necessary to obtain
$K=1$, and we still have $\lim_{N\rightarrow \infty}B_{u+v}=0$,
which means that also the fraction of solved tags is
asymptotically zero. Therefore, in (\ref{efficiency}) we have $ \lim_{N\rightarrow \infty} N^{''}/N=1$, and by (\ref{efficiencyk}) $\lim_{N \rightarrow \infty} L^{''}(N)
/N=0$, so that (\ref{efficiency}) yields $\eta=e^{-1}$.

\section{Practical Issues}\label{sec:practical}
\subsection{Overall Optimization}

\begin{figure}
\centering \includegraphics[width=\columnwidth]{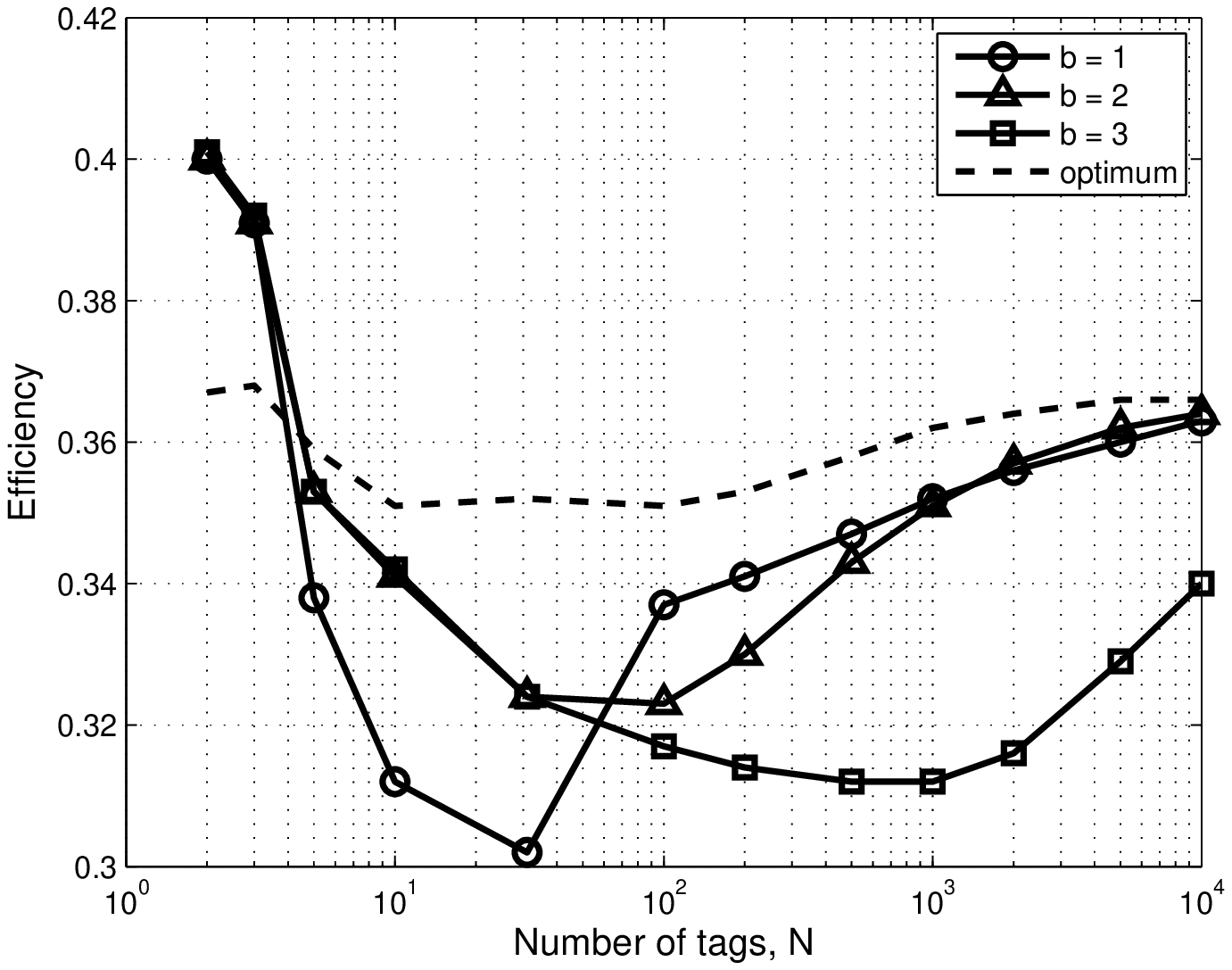}
\caption{Efficiency of AE$^2$ and of overall optimization algorithms versus the tag population
size $N$ for different values of  parameter
$b$.} \label{fig11}
\end{figure}

The analysis carried out in the previous section shows that AE$^2$ is asymptotically  efficient
whatever the values of parameters in (\ref{eq:pre4}) are. Here we investigate the efficiency when
$N$ is finite, in the range $[1,10000]$. Figure \ref{fig11} shows the efficiency of the AE$^2$
procedure for three different values of the parameter $b$, versus the number of tags $N$ to be
identified. We see that in all cases the convergence is assured, although with different
performance. In fact, the best convergence is provided by $b=1$, because it reduces the
approaching phase overhead as predicted by (\ref{efficiencyc}). We see, however, that  for values
of $N\le 100$, the performance appears to deteriorate in all cases, because of the small length of
the observed frame $r_i$, which increases the estimation variance.

The above observation shows that laws (\ref{eq:pre4}) and (\ref{eq:pre2}), though asymptotically
optimal, are non-optimal with finite $N$. On the other side, those laws can  be changed in a
largely arbitrary way without affecting the asymptotic efficiency, as long as the former increases
no more than polinomially, and the latter is an increasing function ultimately converging to
(\ref{eq:pre2}). In the following we show a heuristic method to select those laws optimally, i.e.,
in such a way to provide nearly optimum performance for any $N$.

In the sequel we refer to a mechanism that subdivides the procedure into two phases, namely the
approaching phase, that is roughly composed of the approaching phase and the  converging phase
defined in Sec. \ref{sec:analysis}, and the tracking phase. In the approaching phase we adopt,
instead of (\ref{eq:pre4}) and (\ref{eq:pre2}), laws optimally determined for this phase, and
afterward we turn to (\ref{eq:pre4}) and (\ref{eq:pre2}), thus assuring the convergence of $\hat
n$ and the asymptotic optimality. We can not precisely define when the approaching phase finishes,
since $\hat n$ gets close to $N$ but can rarely match it exactly. However, as the estimation
mechanism of AE$^2$ becomes effective when not all collisions are observed in the frame, we
reasonably assume that the approaching phase finishes at the frame that shows the first non
collided slot.

In looking for optimal laws for the approaching phase, we are able to show that the two of them can
be optimized separately. We assume at first that the estimate $\hat n$ (second law) is doubled at
each frame. The performance indexes of the approaching phase are given by its overhead, that
increases as the length of the real frame increases, and the accuracy of the approach of $\hat n$
to $N$, that can be measured by the traffic per slot $N/z_i=N/\hat n_i$, at the end of the
approaching phase.

We now compare two cases, the first where all the frames of the approaching phase obey $r_i=1$, and
the second when $r_i=2$. In both cases the estimate $\hat n$ is doubled at each frame, and the
approaching phase finishes when a non-collided slot is observed. Assuming that the traffic per
slot is Poisson distributed, independently at each slot, an assumption that becomes more accurate
as $N$ increases beyond a few decades, we can easily find the \emph{a posteriori} traffic $s$ that
maximizes the probability of observing a non collided slot preceded by a sequence of all collided
slot. This traffic turns out to be $s_1=1.4$ and $s_2=1.85$ for the two cases considered above.
Simulations with $N=100$ provides $\bar s_1\simeq 1.5$ and $\bar s_2\simeq 2$ with standard
deviations respectively equal to $0.9$ and $1$, while with $N=10$ provides $\bar s_1\simeq 1.4$
and $\bar s_2\simeq 1.85$ with standard deviations respectively equal to $0.6$ and $0.7$. These
results clearly show that the approach of the first case is slightly better, i.e., closer to $1$,
than the second one. On the other side, increasing the frame length beyond one remarkably
increases the overhead, so that the first case is definitely better than the second one. Based on
this argument we may conclude that making
\begin{equation}\label{eq:pre4b}
r_{i}=1, \qquad i \ge 0,  \end{equation} always provides close to best performance.

In general, the estimate update can be represented by the law
\begin{equation}\label{eq:pre2b} \hat n_{i+1}=H'_i \hat n_i. \end{equation}
Repeating the above evaluations  it appears that, as long as sequence $\{H'_i\}$ is close to all
twos, say in the range $[1.5; 2.5]$, (\ref{eq:pre4b}) is still optimal. We now look for the
optimal $\{H'_i\}$ within the cited range.  Here, we define the optimum sequence as the one that
yields the highest among the minimum efficiencies observed in the whole range of $N$, so that the
efficiency curve turns out as flat as possible.

The problem simplifies in some ways. First, we observe that a sequence $\{H'_i\}$ composed of all
twos yields good performance above $N=1000$, while the minimum of the efficiency appears to be in
the range $ N <1000$. In this range, doubling $\hat n$ at each frame, provides an approaching
phase length that in the majority of cases lasts about $10-12$ frames. This means that our optimum
search can be limited to such a finite sequence. By further  discretizing $H_i$ in steps of $0.1$,
we have been able to perform  an exhaustive search that yields as optimal multipliers the sequence
\begin{equation}\label{eq:pre10} H=[ 2\ \ 2\ \ 2\ \ 2\ \ 1.8\ \ 1.7\ \ 1.7\ \
\ldots, 1.7\ \ \ldots ]. \end{equation} We have  prolonged the above sequence beyond $i=12$ since
this prolongation little, if at all, affects the results with high $N$. The corresponding
efficiency curve is  shown in Fig. \ref{fig11} and positively compares with the curves  seen
before. Here, we can appreciate that the efficiency is always above $0.35$, reaching $e^{-1}$
asymptotically.

\subsection{Implementation issues}

As we have already stated at the beginning  of Sec. \ref{sec:new}, real and virtual frames in
AE$^2$ are compatible with the EPC standard set of commands. The procedure summarized in
(\ref{eq:pre2}), (\ref{eq:pre4}), (\ref{eq:pre4b}) and (\ref{eq:pre10}) can easily be performed in
the reader. However, in this standard, the frame size (virtual, in our case), conveyed by the
reader to tags, must be of the type $2^Q$, with $Q$ in the range from one to $16$. This constraint
impairs the performance even if $N$ is known, since it prevents the optimal assignment $r=n$.

Here we expose an approximate argument that is able to capture the asymptotic
impairment due to the
replacement of the optimal frame length $n$ with the value  closest to $2^Q$,
when the backlog $n$
is known.

Denoting by  $\{n_i\}$  and $\{r_i\}$ the sequence of backlogs and frame
lengths, we assume that
the random variable $n_i/r_i$, i.e., the average number of tags per slot in frame
$i$, presents the
same statistics  regardless of the index $i$ of the frame, which can be
considered a good
approximation for a large part of frames when $N$ is very large. This leads to a
frame efficiency
$\eta_f$ equal for all frames and, therefore, the identification period length
$L(n)$, starting
with $n$ tags, can then be written as $L(n)=\sum_i r_i= n + n(1-\eta_f) + n(1-\eta_f)^2 + \ldots= n/\eta_f$, where, as usual when $n$ is large, we have considered the random variables equal
to their averages. This argument shows that the overall efficiency is $\eta=\eta_f$. In order to
determine $\eta_f$, we
find the statistics of $n_i/r_i$, when the power index $Q_i$ of the frame length
is assumed such
that $2^{Q_{i}}$ is the closest to $n_i$. To the purpose, let assume that such
index is $Q_i=q$. This
means that $n=n_i$ is in the range $[2^q-2^{q-2}, 2^q+2^{q-1}]$, being with
equal probability
higher or lower than $2^q$. Similarly, $s=n/r=n/2^q$ has a distribution that is
bivariate, namely
uniform within $[3/4, 1]$ with probability $1/2$, and  uniform within $[1,3/2]$
again with
probability $1/2$. The efficiency, or probability of a slot being successful,
can be expressed as $\left(1-1/r\right)^{n-1}= \left(1-s/n\right)^{n-1}
\rightarrow e^{-s}$ and,  averaged over the  distribution of $s$ already cited, provides
$\eta=\eta_f= 0.3562$.

\begin{figure}
\centering \includegraphics[width=\columnwidth]{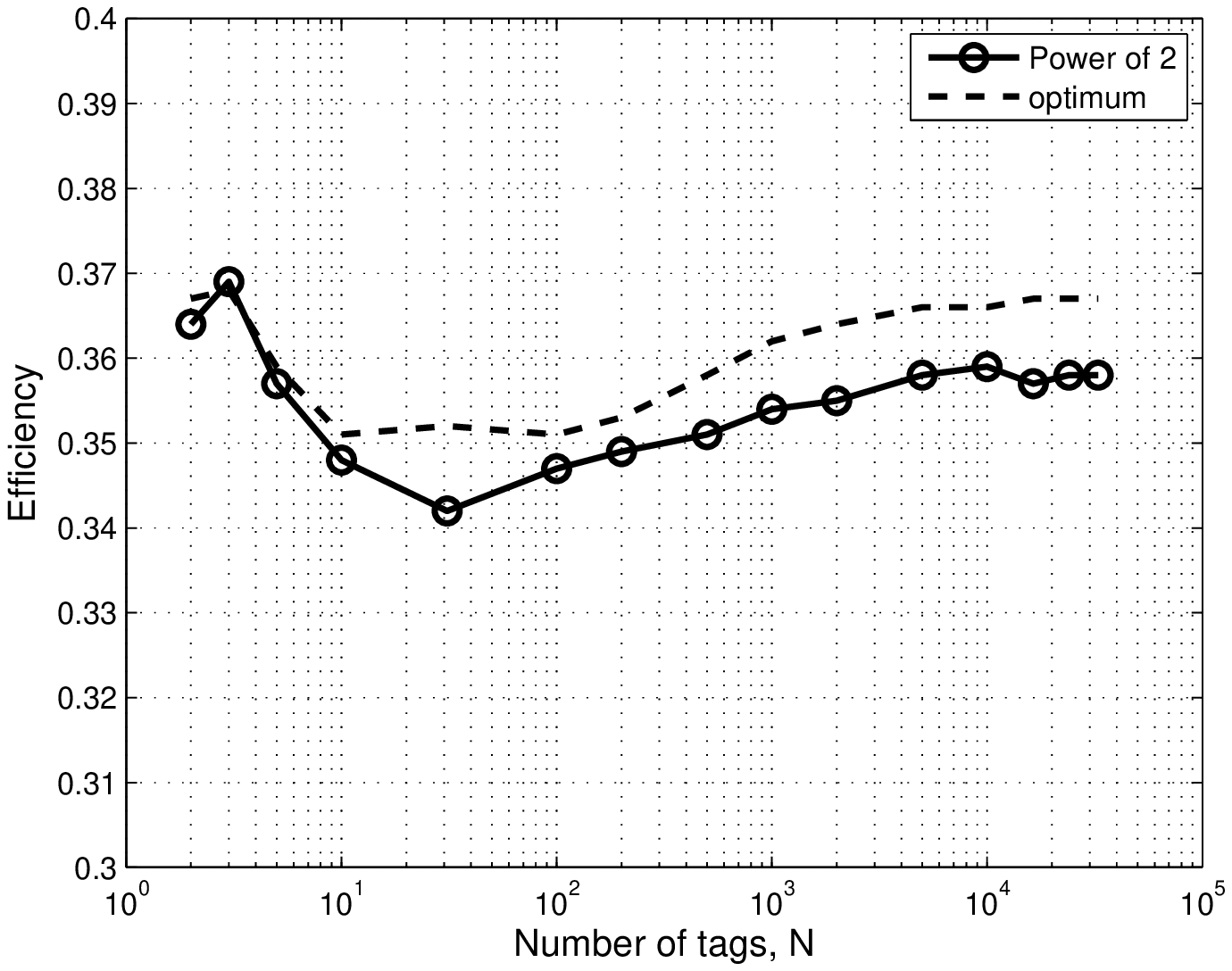}
\caption{Efficiency of AE$^2$ and overall optimization algorithms versus the tag population
size $N$ for different values of  parameter
$b$.} \label{fig13}
\end{figure}

The above value of $\eta$ almost perfectly matches  the asymptotic values we
have found by
simulation, namely $0.357$. Figure \ref{fig13} shows the efficiencies of AE$^2$
and its
constrained version. In the latter,  sequence (\ref{eq:pre10}) has been replaced
by a sequence of
all $2$, whereas (\ref{eq:pre2}) is replaced by the closest $2^Q$. The results
clearly show the
impairment due to the constraint, but they also show that, also in the constrained
version, our
procedure is able to asymptotically reach the theoretical performance.

\section{Conclusions}
\label{sec:conclusions} In this paper we have presented an asymptotic analysis
of Schoute's
backlog estimate for DFA, applied to the RFID environment. The
analysis shows that
the asymptotic efficiency of this estimate is $0.311$, far less than the
theoretical maximum
$e^{-1}\approx 0.368$, achieved when the backlog is known. The analysis shows
that the performance loss is due to the slow convergence of the estimate and suggests how this
impairment can be
avoided. Using these results we have introduced the AE$^2$, a  procedure that
asymptotically
reaches the theoretical maximum, and that can be set to achieve an efficiency
close to maximum for
any finite value of the population size.  We have also derived the asymptotic
efficiency when the frame
size is constrained to be a power of two, as required by RFID standards for
DFA,
with which the proposed protocol is fully compatible.  Although the gain with
existing mechanisms
is moderate (in the $10-20 \%$ range), we remark that the proposed procedure is
able to cope with
any tag number, and does not present an hard limit on the maximum number of tags
to be resolved.

%\appendices
\section*{Appendix A}\label{sec:appendix_A}

If $n_i$ and $r_i$ are both large, collided slots in frame $i$ become
distributed according to a
binomial distribution with average $r_i p_c $ and variance $r_i p_c (1-p_c)$,
which is upper
bounded by $r_i$. Therefore, being,  in the approaching phase and for large $r$,
 $r_{i+1} = H
c_i$  we have
\begin{equation}\label{eq:var00}
 \mbox{VAR}\{r_{i+1}|n_i,r_i\}=H^2\ \mbox{VAR}\{c_i|n_i,r_i\}\le H^2 r_i.
 \end{equation}
 Since the number of collisions can not be larger than $N/2$ it follows that
\begin{equation}\label{eq:var00x}
r_{i}\le H  \frac{N}{2},
 \end{equation}
 for all $i$.
Substituting (\ref{eq:var00x}) into (\ref{eq:var00}) yields
 $ \mbox{VAR}\{r_{i+1}|n_i,r_i\}\le N d$, where $d$ is a constant value.  The Chebyshev's inequality used with the above
bound yields
\begin{equation*}%\label{eq:var02}
\Pr(|r_{i+1}-{\cal R}_{i+1}|\ge\epsilon N|n_i,r_i)\le \frac{d}{N\epsilon^2},
 \end{equation*}
 that can be reduced to
 \begin{equation}\label{eq:var03}
 \Pr(|r_{i+1}-{\cal R}_{i+1}|\ge\epsilon N)\le \frac{d}{N\epsilon^2}.
 \end{equation}
 Relation (\ref{eq:var03})
 shows that,  for $N \rightarrow \infty$, we have $ r_i/N\rightarrow  {\cal
R}_i/N, $  where the convergence is in probability.
Much in the same manner one can show that  $ n_i/N\rightarrow  {\cal N}_i/N, $
 and, therefore, we have  $ n_i/r_i\rightarrow  {\cal N}_i/{\cal R}_i $ in
probability.

\section*{Appendix B}\label{sec:appendix_B}
Here we consider sequence  $\{\mathcal{R}_i\}$ during the first phase, where all
the slots are
collided, i.e., $\mathcal{C}_i=\mathcal{R}_i$ and relation (\ref{eq:sl1a})
becomes $\mathcal{R}_{i+1}=\expop\{Hc_i+\xi_i\}=H\mathcal{C}_i+\Xi_i=H\mathcal{R}_i+\Xi_i$, for $i \ge 0$.
On the other side we have $R_{i+1}=HR_i$, for $i \ge 0$,
with $\mathcal{R}_0=R_{0}=r$. Solving the recursions we get
\begin{equation}\label{eq:aa5}
\mathcal{R}_i=r H^{i}+ \sum_{k=0}^{i-1} H^{i-1-k}\Xi_k
\end{equation}
\begin{equation}\label{eq:aa6}
 R_i=r H^{i},
\end{equation}
for $i\ge 0$. Relation (\ref{eq:aa5}) can be rewritten as
$\mathcal{R}_i=R_i+ \sum_{k=0}^{i-1} H^{i-1-k}\Xi_k$.
Since $ |\Xi_k|\le 0.5<1$, and being $\sum_{k=0}^{i-1} H^{k}= (H^{i}-1)/(H-1)$, we can write
\begin{equation*}%\label{eq:aa7}
R_i -\frac{H^{i}-1}{H-1} < \mathcal{R}_i < R_i +\frac{H^{i}-1}{H-1} , \ \ \ \ \
\ i \ge 0,
\end{equation*}
and
\begin{equation*}%\label{eq:aa8}
1- \frac{f(H)}{r(H-1)} <
\frac{\sum_{i=0}^{\infty}\mathcal{R}_i}{\sum_{i=0}^{\infty}R_i}<1+
\frac{f(H)}{r(H-1)},
\end{equation*}
with $f(H)=(\sum_{i=0}^{\infty}(H^{i}-1))/(\sum_{i=0}^{\infty}H^i)$, having exploited (\ref{eq:aa6}). Since it is $H^{i}-1 < H^i$, for $i \ge 0 $, we
also have $f(H)<
1$, and finally $\lim_{r\rightarrow\infty} (\sum_{i=0}^{\infty}\mathcal{R}_i)/(\sum_{i=0}^{
\infty}R_i)=1$.

\baselineskip 10pt

\bibliography{biblio}
\bibliographystyle{IEEEtran}

% %\newpage
% %\begin{table*}[t!]
% %\caption{Values of throughput $n/L(n, r)$ versus $n$ for Dynamic Frame Aloha,
% % when $n$ is estimated with Schoute's method. The two right-most columns have
% been evaluated by
% %computer simulation.} \hspace*{-0.18cm} \centerline{
% %\begin{tabular}{|c|ccccccc|cc|}
%  % \hline & $N=2$ & $N=3$ & $N=5$ & $N=10$ & $N=20$ & $N=30$ & $N=100$ & $N=500$
% & $N=1000$\\ \hline
% %$N/L^*(N )$ & 0.500 & 0.471 &0.441 & 0.413 &0.395 &0.387 &0.375 &0.370 &0.369
% \\
% % $N/L(N, N)$ & 0.500 &0.468 &0.434 & 0.407 &0.391 &0.385 &0.374 &0.369 &0.369
% \\
% % $N/L(N, 1)$ & 0.400 & 0.391 & 0.353 & 0.342 & 0.330 & 0.324 &0.317 & 0.312 &
% 0.312 \\
% % $N/L(N, 10)$ & 0.192 &0.269 &0.367 & 0.407 &0.368 &0.349 &0.323 &0.314 &0.312
% \\
% % $N/L(N, 100)$ & 0.020 &0.030 &0.050 & 0.098&0.185 &0.253 &0.374 &0.327 &0.319
% \\
% %  \hline
% %\end{tabular}}
% %\label{tab1}
% %\end{table*}

\newpage

\begin{table}[h]
\caption{Analytical values of the asymptotic efficiency of DFA with
Schoute's estimate for
different values of the parameter $K_u$.} \centerline{
\begin{tabular}{|c|ccccccc|}
 \hline & $K_u=20$ & $K_u=25$ & $K_u=30$ & $K_u=35$ & $K_u=40$ & $K_u=45$ &
$K_u=47.8$ \\ \hline
  $N/L(N)$ &0.31125  &0.31127&0.31125& 0.31122 & 0.31122  &0.31123 & 0.31125 \\
  \hline
\end{tabular}}
\label{tabfr4b}
\end{table}

\newpage

\newpage

\newpage

% %\newpage
% %
% %\begin{table*}[t!]
% %\caption{\it Throughput of the Q-Algorithm under different tags numbers $N$ and
% initial frame
% %lengths ($z_0$).} \centerline{
% %\begin{tabular}{|c|cccccccc|}
% %  % after \\: \hline or \cline{col1-col2} \cline{col3-col4} ...
% %  \hline $N/L_N$  & $N=2$ & $N=3$ & $N=5$ & $N=10$  & $N=20$ & $N=30$
% &$N=1000$  &$N=10000$ \\ \hline
% %             $z_0=1$  & 0.344 &  0.314 & 0.304 & 0.302& 0.312 &0.318 & 0.354 &
% 0.359 \\
% %             $z_0=16$  &0.144 & 0.199 & 0.275 & 0.345& 0.358&0.355 & 0.355  &
% 0.359  \\
% %    \hline
% %\end{tabular}}
% %\label{tabq11}
% %\end{table*}

\newpage

\newpage

\newpage

\newpage

\newpage

\newpage

\end{document}